\documentclass{article}

\ifx\directlua\undefined\ifx\XeTeXcharclass\undefined
  \usepackage[utf8]{inputenc}                           
  \else\RequirePackage[no-math]{fontspec}[2017/03/31]\fi 
  \else\RequirePackage[no-math]{fontspec}[2017/03/31]\fi 
\usepackage[sort&compress,square,numbers]{natbib}
\usepackage{xcolor}
\usepackage{url}
\usepackage[dvipsnames]{xcolor}
\usepackage{fix-cm}  
 \usepackage{graphicx}
  \usepackage{multirow}
 \usepackage{amsmath,amsfonts, amssymb}
 \usepackage[margin=1in]{geometry}
 
\author{%
Ergun Simsek\textsuperscript{1,*},
Shao-Chien Ou\textsuperscript{2,3},
Gregory Moille\textsuperscript{2,3},
Kartik Srinivasan\textsuperscript{2,3}
}
 
\begin{document}
\title{Microring Resonator Dispersion Metrology with Neural Networks}

\maketitle
\begin{center}
\textsuperscript{1}University of Maryland, Baltimore County, Baltimore, USA \\[2mm]
\textsuperscript{2}Joint Quantum Institute, NIST/University of Maryland, College Park  \\[2mm]
\textsuperscript{3}National Institute of Standards and Technology, Gaithersburg, USA \\[2mm]
\textsuperscript{*}Corresponding author's e-mail address: simsek@umbc.edu \\
\end{center}

\begin{abstract}
Precise knowledge of resonator dispersion, from both geometric and material contributions, is essential for reliable high-performance nonlinear integrated photonics devices, such as optical parametric oscillators, frequency doublers, and integrated optical frequency combs. However, direct measurements at the fabrication level provide limited knowledge, whether through destructive cross-section imaging or non-destructive ellipsometry, while complete optical characterization that enables precise dispersion metrology is time-consuming and poorly suited for mass-scale foundry fabrication.
In this work, we develop a machine learning framework to solve three complementary problems: (i) predicting resonator geometric dimensions, (ii) identifying the correct material dispersion, and last, but not least, (iii) precisely reconstructing the integrated dispersion spectrum directly from ring dimensions. These three neural networks together enable both inverse and forward characterization of microring resonators. Using numerically generated datasets based on Sellmeier-type material models, we demonstrate $<1$~nm ring dimension prediction accuracy without noise, $<8$~nm prediction accuracy with $\approx45$ dispersion samples under a realistic frequency measurement noise level ($\pm~50$~MHz), and $\approx$16~nm prediction accuracy at a higher noise level ($\pm~200$~MHz). The Sellmeier model classification exceeds 99\% accuracy in all cases. Importantly, dispersion sampled far from the pump resonances proves most informative, reducing full-spectrum characterization requirements. The forward-prediction network reconstructs dispersion spectra from the ring dimensions with high accuracy, providing a rapid alternative to numerical simulations and enabling quick assessment of the dispersion regime. By combining both inverse and forward predictions, our results highlight the potential of machine learning applied to dispersion data as a rapid, non-destructive tool for wafer-scale quality control and process monitoring in photonic foundries.
\end{abstract}

\section{Introduction}
\renewcommand{\arraystretch}{1.2}
\begin{figure*}[!t]
    \centering
    \includegraphics[width=0.85\linewidth]
    {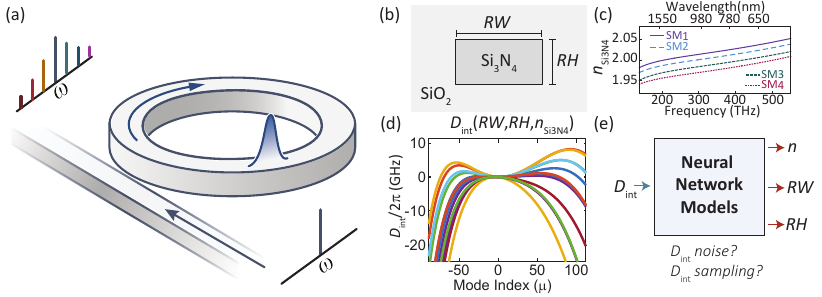}
    \caption{(a) Dissipative Kerr soliton (DKS) microcombs convert a single frequency laser to a frequency comb, often in Si$_3$N$_4$ microring resonators. The properties of the frequency comb depend on the dispersion of the microring resonator, quantified by the integrated dispersion $D_\text{int}$, which describes the deviation of the microring mode frequencies from a uniformly spaced grid.  $D_\text{int}$ depends on (b) the resonator geometry, and in particular, the ring width ($RW$) and ring height ($RH$) and (c) the refractive index dispersion of the constituent materials. For Si$_3$N$_4$, its dispersion will be impacted by different gas precursor ratios in the growth process (labeled SM1 to SM4 here, and described in further detail later). (d) Example simulated $D_\text{int}$ curves for different geometric and material parameters. (e) Here, we use neural networks to understand how well we can extract geometric and material parameters given $D_\text{int}$ data. We further consider how well this extraction works in the presence of noise on the $D_\text{int}$ data associated with measurement uncertainty, and what level and types of downsampling of $D_\text{int}$ data can still lead to good parameter extraction.}
    \label{fig:Intro}
\end{figure*}

The miniaturization and integration of optical frequency combs promises revolutionary advances in telecommunications \cite{ucombs_for_comm}, precision metrology \cite{hansch2}, and quantum information science \cite{ucombs_for_quantum}. Among various approaches, optical frequency comb generation using the Kerr nonlinearity in microresonators (microcombs) has emerged as particularly promising due to its compact footprint, low power consumption, and potential for large-scale integration \cite{DelHaye:2007, Kippenberg:2018,gaeta_photonic-chip-based_2019,chang_integrated_2022}. In particular, silicon nitride (Si$_3$N$_4$) microring resonators have emerged as a platform of choice because of their low propagation loss, high Kerr nonlinearity, and compatibility with standard CMOS fabrication processes \cite{Moss:2013, pfeiffer_photonic_2016, ji_methods_2021, simsek2025stop}. These properties, together with broadband dispersion engineering capability~\cite{okawachi_octave-spanning_2011}, have led to the demonstration of octave-spanning dissipative Kerr soliton (DKS) combs directly on a chip~\cite{li_stably_2017,pfeiffer_octave-spanning_2017,yu_tuning_2019,weng_dual-mode_2022,wu_vernier_2023,zang_foundry_2024} (Fig.~\ref{fig:Intro}(a)). Such octave-spanning DKS microcombs can be self-referenced to stabilize their carrier-envelope offset frequencies~\cite{spencer_optical-frequency_2018,newman_architecture_2019,moille_all-optical_2025}, opening the door for many applications in optical frequency synthesis, optical atomic clocks, ultra-low noise microwave generation, and spectroscopy.

For a usable octave-spanning DKS microcomb, the resonator dispersion is typically tailored to enable phase matching between the DKS and the microring resonance. This phase matching generates dispersive waves (DWs)~\cite{brasch_photonic_2016}, engineered to be at quasi-harmonic frequencies~\cite{briles_interlocking_2018}, which enhance the power of phase-matched comb teeth and drastically improve microcomb self-referencing~\cite{spencer_optical-frequency_2018,moille_all-optical_2025}.
The resonator dispersion has two contributions, a geometric component that is largely determined by the ring cross-section (Fig.~\ref{fig:Intro}(b)), and a material component that comprises both core and cladding contributions (Fig.~\ref{fig:Intro}(c)). The fabrication precision directly impacts the geometric dispersion, as even $\lesssim10$~nm  changes in ring width~\cite{briles_interlocking_2018,yu_tuning_2019} and thickness~\cite{moille_tailoring_2021} can modify the comb bandwidth and shift dispersive wave frequencies by $>5$~THz. The material growth directly determines the material dispersion, which for amorphous materials such as Si\textsubscript{3}N\textsubscript{4} is sensitive to the growth parameters such as the gas settings~\cite{Moille:21}. 

Methods for in-line nondestructive monitoring remain limited to ellipsometry measurements on a restricted set of test wafers, and exclude geometrical dispersion contributions. To monitor this second contribution, destructive analysis is necessary to cross-section a given set of fabricated microring resonators. Thus, such methods can be insufficient for optimization and monitoring of large-scale fabrication processes, such as those at the American Institute for Manufacturing Integrated Photonics (AIM Photonics), where substantial enough variations can occur across a single wafer \cite{fahrenkopf2019aim} to strongly impact microcomb performance~\cite{SO2025}. An alternative, more direct, and non-destructive approach characterizes the microring resonator optically by measuring the residual dispersion of mode $\mu$ (Fig.~\ref{fig:Intro}(d)), also called the integrated dispersion ($D_\mathrm{int}(\mu)$), against a fixed grid of frequency markers. However, complete characterization of a given resonator is time-consuming, may require a large set of tunable lasers, and is therefore unsuitable for rapid testing of many fabricated devices for dispersion calibration across a wafer. 

Recent work has highlighted the promise of machine learning (ML) and inverse-design methods for addressing such topics. For example, ML-based inverse design has been successfully applied to waveguide dispersion engineering, demonstrating agile dispersion profiles for broadband and nonlinear applications \cite{Wang2021}. Similarly, random forest and decision tree regressors have been used to infer resonator geometries directly from dispersion measurements, with experimental verification on integrated Si$_3$N$_4$ platforms \cite{Pal:23}. In ref.~\cite{ES_Attention:2023}, Soroush \textit{et al.} showed that broadband resonator-waveguide coupling coefficients can be predicted from a small set of measured (or calculated) coupling coefficient values. More generally, automated inverse design of on-chip resonators has been demonstrated, overcoming long-standing challenges associated with the highly nonconvex optimization landscape of resonant devices \cite{Ahn2022}. Together, these advances indicate that ML provides a powerful route for linking measurable dispersion signatures to underlying device properties, enabling both design automation and device characterization.

In this work (Fig.~\ref{fig:Intro}(e)), we build upon these concepts and address three complementary problems using neural networks trained on numerically generated $D_{\mathrm{int}}(\mu)$ datasets and ring dimensions. First, we demonstrate inverse prediction of microring geometry: given integrated dispersion measurements, our regression model infers the ring height and width with $<1$~nm accuracy in noiseless conditions. Second, we classify the material dispersion model by training a neural network to discriminate between different Sellmeier-type refractive index models with an accuracy exceeding 99\%. This classification task may be particularly important for identifying variations in deposition processes that alter the material response. Third, we present a forward-prediction neural network that estimates the coefficients of a sixth-order polynomial to reconstruct the integrated dispersion spectrum directly from the ring dimensions. This forward model provides rapid evaluation of $D_{\mathrm{int}}$ without repeated numerical simulations and establishes a bidirectional ML framework linking geometry, material models, and spectral signatures. In other words, this forward model can be used as a predictive tool to determine whether a planned resonator design will yield a $D_\text{int}$ spectrum consistent with the desired dispersion regime. In this way, our framework not only enables robust geometry and material classification from spectral data but also provides a direct route to evaluating and optimizing dispersion profiles during the design stage.  

We further investigate the robustness of our models under realistic measurement noise and analyze the trade-off between accuracy and the number of $D_{\mathrm{int}}$ samples used. Our results highlight two key findings: (i) in the presence of realistic measurement uncertainty of individual resonance frequencies of approximately $\pm 50$~MHz, $<$8~nm accuracy can be obtained using as few as 45 $D_{\mathrm{int}}$ samples, provided that the samples are chosen sufficiently far from the pump frequency, and (ii) for less precise systems with noise levels on the order of $\pm 200$~MHz, the expected errors approximately double, reaching $\approx$16~nm. Our results provide practical guidelines for experimental implementations by linking measurement system precision to the number of required $D_{\mathrm{int}}$ samples and the achievable accuracy. 

\section{Integrated Dispersion Calculations\label{DintCalculation}}
We consider a microring resonator composed of Si$_3$N$_4$ as the core material and SiO$_2$ as the cladding. The outer ring radius is fixed at $23~\mu$m, while the ring height and width are varied to construct a comprehensive dataset for dispersion analysis. Specifically, the height is swept from $620$~nm to $720$~nm and the width is swept from $750$~nm to $950$~nm in increments of $10$~nm, resulting in $441$ unique training designs. Additionally, we generate a test dataset by randomly selecting 50 distinct integer width–height combinations (in nanometers) from the same ranges as the training dataset.

To accurately capture material dispersion, we employ wavelength-dependent relative permittivity models for both Si$_3$N$_4$ and SiO$_2$. The relative permittivity of Si$_3$N$_4$ under different gas ratios is modeled as
\begin{equation}
\varepsilon_r^{\text{Si}_3\text{N}_4}(\lambda) = 1 + \frac{A}{1 - (B/\lambda)^2} - C\lambda^2,
\end{equation}
where $\lambda$ is the wavelength in $\mu$m and the parameters $(A,B,C)$ depend on the gas ratio used during deposition \cite{Moille:21}. The parameter sets are summarized in Table~\ref{tab:si3n4_disp}.
\begin{table}[!h]
\footnotesize
\centering
\caption{Dispersion model parameters for $\varepsilon_r$ of Si$_3$N$_4$ at different gas ratios \cite{Moille:21}.} 
\label{tab:si3n4_disp}
\begin{tabular}{ccccc}
\hline
Model  & Gas ratio  & $A$     & $B$ & $C$ \\
Name & (NH$_3$:SiH$_2$Cl$_2$) & \ &  \ & \ \\
\hline
SM$_1$ & 3:1   & 3.025 & 0.13534 & 0.0230 \\
\hline
SM$_2$ & 5:1   & 2.973 & 0.13475 & 0.0220 \\
\hline
SM$_3$ & 7:1   & 2.883 & 0.13364 & 0.0244 \\
\hline
SM$_4$ & 15:1  & 2.842 & 0.14018 & 0.0181 \\
\hline
\end{tabular}
\end{table}
\normalsize

For the surrounding SiO$_2$, we adopt the Sellmeier-type relation~\cite{TAN1998158}
\begin{equation}
\varepsilon_{r}^{\text{SiO}_2}(\lambda)
= 1 + \sum_{i=1}^3 \frac{A_i}{1 - \left(\tfrac{B_i}{\lambda}\right)^{2}}.
\end{equation}
The coefficients are given in Table \ref{tab:sio2}.

\begin{table}[!h]
\footnotesize
\centering
\caption{Dispersion model parameters for SiO$_2$.} 
\label{tab:sio2}
\begin{tabular}{ccc}
\hline
i & $A_i$ & $B_i$ \\
\hline
1 & 0.6961663 & 0.0684043 \\
\hline
2 & 0.4079426 & 0.1162414 \\
\hline
3 & 0.8974794 & 9.896161 \\
\hline
\end{tabular}
\end{table}
\normalsize

With these permittivity models, the calculation of $D_\text{int}$ proceeds as follows. We first compute the effective modal indices, $n_{\text{eff}}(\lambda)$, using our eigenmode solver \cite{Die_Ring_Solver, simsek2025stop} at $171$ discrete wavelength values uniformly distributed between $750$~nm and $1.6~\mu$m for the training and test datasets. The propagation constant is given by $\beta(\lambda) = 2\pi n_{\text{eff}}(\lambda)/\lambda$, and the azimuthal mode number $m$ satisfies the resonance condition $m = \beta R_{\text{c}}$, where $R_{\text{c}}$ is the central radius of the microring. Since the values obtained directly from $\beta R_{\text{c}}$ are generally non-integer, we first fit a fifth-order polynomial to the $n_{\text{eff}}$ vs $m$ relationship and then evaluate the effective index at integer mode numbers $m_i$ within the physically relevant range. The resonance frequencies are then computed as:
\begin{equation}
f_i = \frac{c m_i}{n_{\text{eff},i} L},
\end{equation}
where $L = 2\pi R_{\text{c}}$ is the ring perimeter and $c$ is the speed of light.

We define the pump mode as the resonance closest to the target pump wavelength of 1060 nm. The free spectral range $D_1$ at the pump frequency can be calculated from $D_1 = f_1 - f_0$ or from $D_1 = (f_1 - f_{-1})/2$,  where $f_0$ is the pump resonance frequency and $f_{-1}$ and $f_1$ are the resonance frequency of the adjacent modes. Then, the integrated dispersion is evaluated as a function of relative mode number $\mu = m_i - m_0$, where $m_0$ is the pump mode number:
\begin{equation}
D_{\mathrm{int}}(\mu) = f_\mu - (f_0 + D_1 \mu),
\end{equation}
where $f_\mu$ is the resonance frequency corresponding to relative mode number $\mu$. This procedure yields 201 integrated dispersion values at integer relative mode numbers, determined from the original 171 wavelength points. While our primary analysis focuses on these $D_{\mathrm{int}}$ profiles, we later incorporate $D_1$ as a supplementary feature to account for its influence on ring dimension predictions and to evaluate the model's robustness against frequency offsets.

\begin{figure}[!h]
    \centering
    \includegraphics[width=0.5\linewidth]{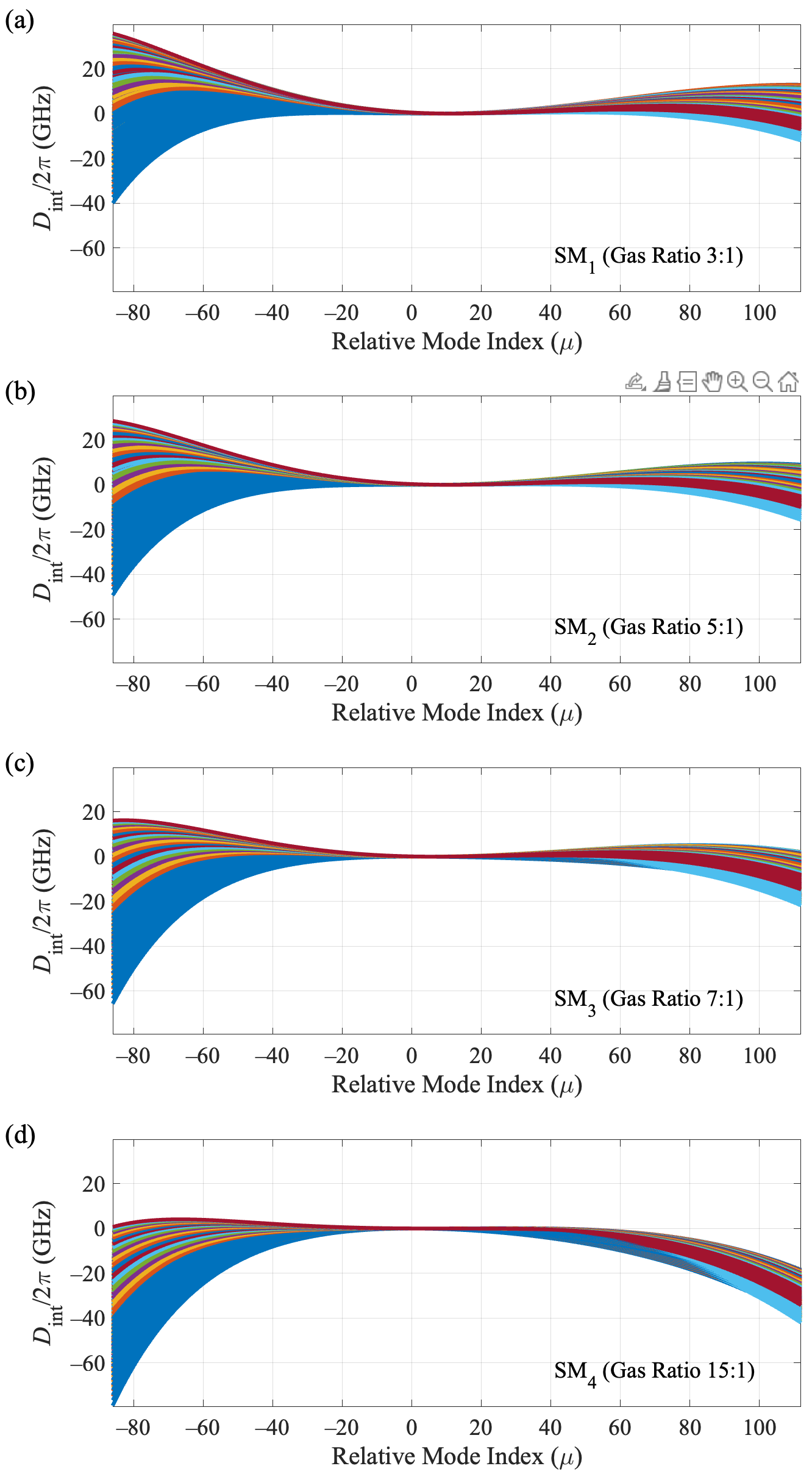}
    \caption{Integrated dispersion curves ($D_{\rm{int}}/2\pi$) as a function of mode index for Si$_3$N$_4$ films grown with the four precursor gas ratios outlined in Table~\ref{tab:si3n4_disp}: (a) 3:1, (b) 5:1, (c) 7:1, and (d) 15:1. Each sub-panel contains 441 distinct $D_{\rm{int}}/2\pi$ curves that span a thickness variation between 620~nm and 720~nm (10~nm step size) and ring width variation between 750~nm and 950~nm (10~nm step size).}
    \label{fig:Dint_SMs}
\end{figure}

Figure \ref{fig:Dint_SMs} shows the integrated dispersion curves as a function of mode number for different precursor gas ratios of 3:1 (SM$_1$), 5:1 (SM$_2$), 7:1 (SM$_3$), and 15:1 (SM$_4$). At low gas ratios, the dispersion curves spread out more significantly, especially at large negative mode indices. This means that changes in geometry (ring width/height) lead to large and distinguishable changes in $D_{\rm{int}}$. With increasing gas ratios, the curves become much more   ``compressed.'' Many different geometries yield very similar $D_{\rm{int}}$ profiles, especially near the pump (mode index 0). This loss of distinguishability implies that the input features, namely the $D_{\mathrm{int}}$ curves, carry less unique information about the underlying geometry. Consequently, while the neural network can still perform reliably on noiseless synthetic datasets where subtle differences remain discernible, its predictive accuracy is expected to degrade in experimental conditions where random noise is unavoidable. 

\section{Numerical Results}
\begin{figure}[!h]
    \centering
    \includegraphics[width=0.4\linewidth]{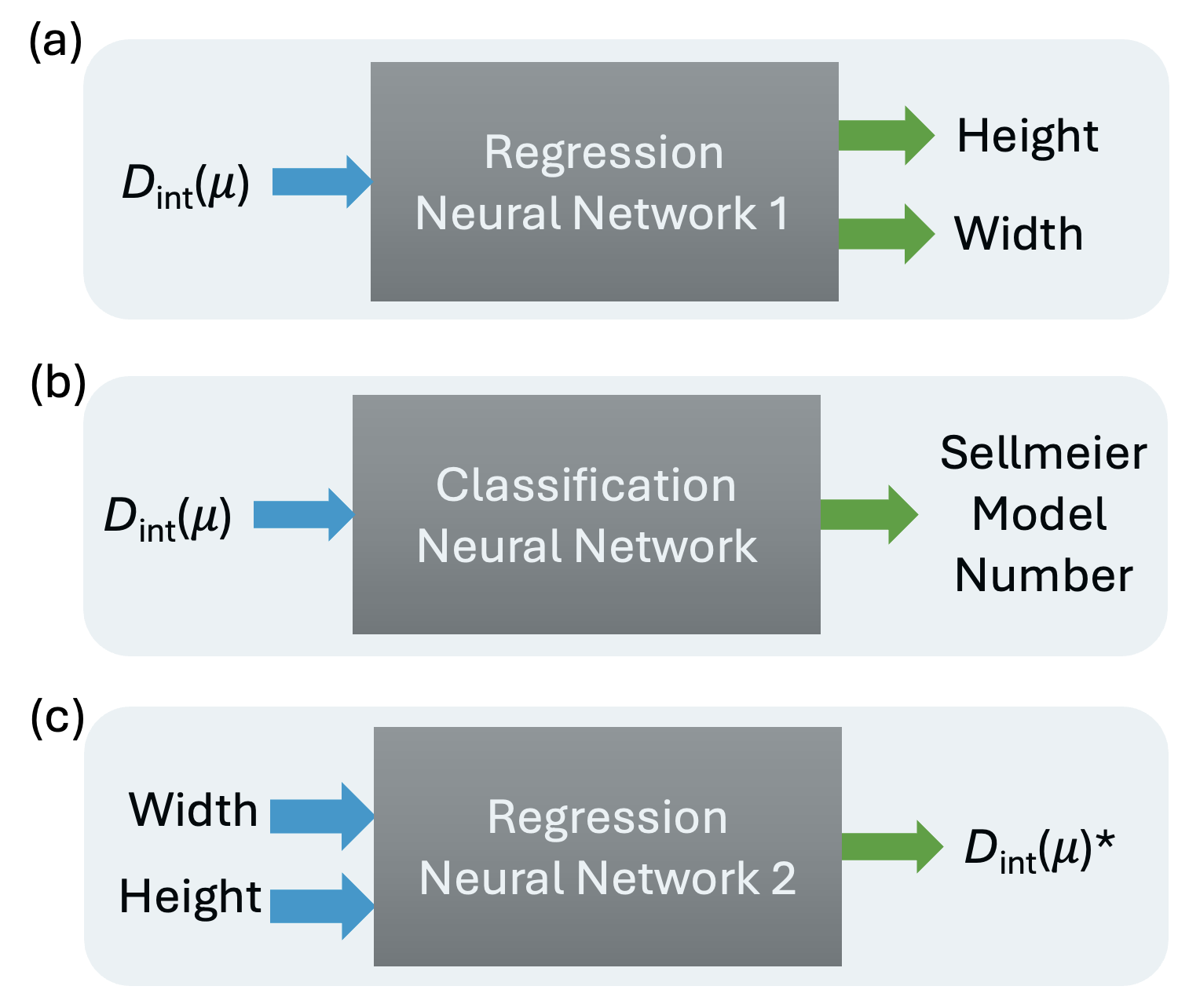}
    \caption{Schematic diagrams of the three neural network architectures used in this study: (a) Regression model for predicting the ring width and height from the integrated dispersion ($D_\text{int}(\mu)$). (b) Classification model for identifying the  Sellmeier model from dispersion data. (c) Another regression model for predicting the coefficients of a $6^{\rm{th}}$ order polynomial fitting the integrated dispersions from the ring widths and heights. To emphasize that we regress on the polynomial coefficients, not on the integrated dispersion directly, we use an asterisk. }
    \label{fig:networks}
\end{figure}

In this study, we used three distinct neural network architectures to address three problems associated with the integrated dispersion curves of dielectric ring resonators, as illustrated in Fig.~\ref{fig:networks}.

\subsection{Dimension Prediction}
\begin{figure}[!h]
    \centering
    \includegraphics[width=0.5\linewidth]{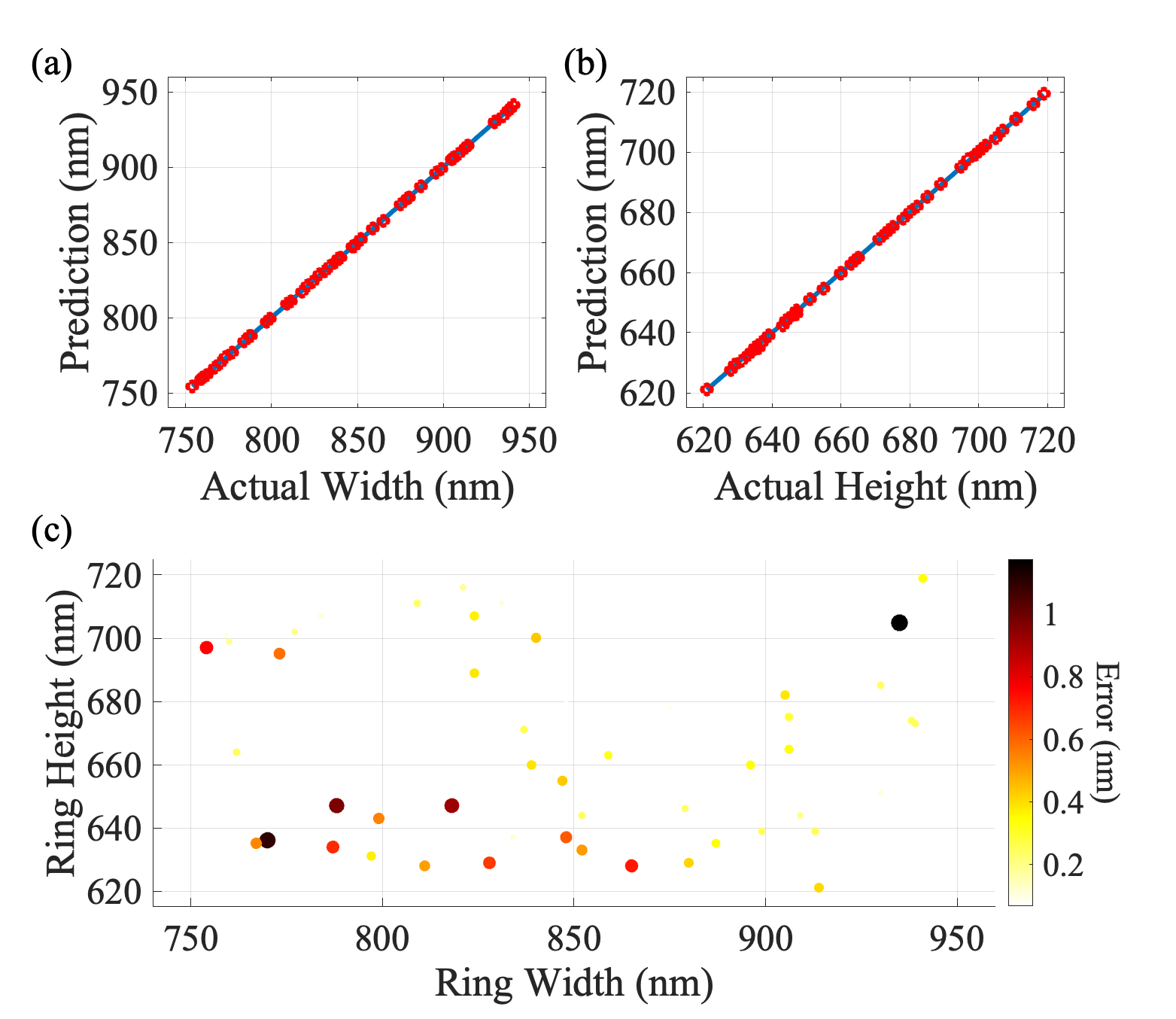}
    \caption{Performance of the regression neural network trained and tested on datasets generated using the Sellmeier model 1. (a) and (b) Predicted versus true values for ring width and ring height, respectively. (c) Error distribution in the actual width–height plane. We calculate the overall error with the formula $\sqrt{(w_a-w_p)^2+(h_a-h_p)^2}$, where $w$ and $h$ represent width and height, and subscripts $a$ and $p$ refer to actual and predicted values.}
    \label{fig:NN3_SM1}
\end{figure}

The first network, Fig.~\ref{fig:networks}(a), is designed for regression and aims to predict the width and height of the resonator from the $D_\text{int}$ data. In this setting, the input consists of the normalized $D_\text{int}$ values determined over 201 mode numbers, while the output is the width and height of the rings. The network consists of 20 fully connected hidden layers, each containing 64 neurons, with hyperbolic tangent (tanh) activation functions applied to introduce nonlinearity. The output layer contains two linear neurons corresponding to the width and height predictions. Training is performed using the Levenberg--Marquardt optimization algorithm \cite{LMoptimization} with a mean squared error loss function, a batch size of 32, and convergence typically achieved within 200 epochs.
To evaluate the accuracy of the regression network in predicting the width and height of the ring resonators, we first train the model and then test it using datasets created with the Sellmeier model~1 (gas ratio of 3:1). The results are summarized in Fig.~\ref{fig:NN3_SM1} as follows.

The panels (a) and (b) of Fig.~\ref{fig:NN3_SM1} display the one-to-one correspondence between predictions and ground truth for ring width and height, respectively. In both cases, the data points are tightly clustered along the diagonal, indicating that the network is able to capture the mapping with remarkable precision. The near-perfect linearity of these plots suggests that the model is not only accurate but also unbiased across the full parameter range of the test set. In panel (c), we plot the error over the actual width–height plane. The circles' radii increase, and their colors get darker, with increasing error. While the vast majority of points exhibit excellent agreement ($<1$~nm error), slightly larger deviations on the order of 1~nm occur in regions where the ring height is small. This is consistent with the fact that, in this regime, variations in height do not significantly alter the integrated dispersion $D_{\rm{int}}$, making the inverse mapping less well-conditioned and more challenging for the network to resolve uniquely.

\begin{figure}[!h]
    \centering
    \includegraphics[width=0.5\linewidth]{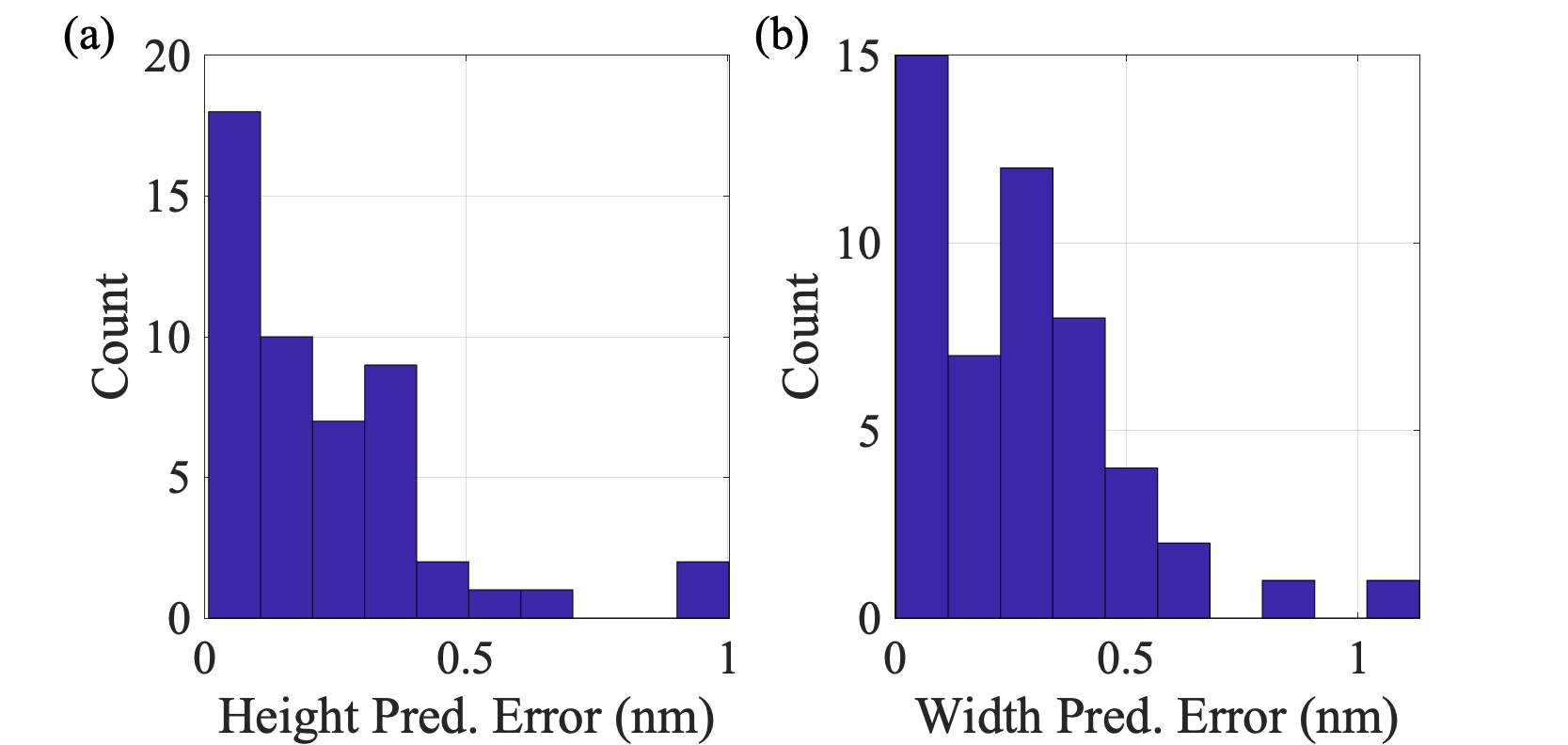}
    \caption{Histograms of the absolute prediction errors for (a) height and (b) width, respectively, for the regression neural network studied in Fig.~\ref{fig:NN3_SM1}. In 49 out of 50 test cases, the errors remain below 1 nm.}
    \label{fig:hist}
\end{figure}

Figure \ref{fig:hist} provides histograms of the absolute errors $|h_a-h_p|$ and $|w_a-w_p|$ for (a) height and (b) width prediction, respectively. These distributions confirm the high accuracy of the network: for 49 out of 50 test cases, the prediction error in both width and height was below 1~nm. We obtained similar results with the datasets created with the other three Sellmeier models. We do not provide them here for the sake of brevity.

To benchmark the performance of our regression network against alternative machine-learning approaches, we also evaluated several standard models—including linear regression, decision trees, random forests, support vector machines, and Gaussian process regression—using the same training and test datasets. The results, summarized in Table~\ref{tab:ML_models}, show that while the Gaussian process model achieves the lowest mean prediction error, its computational cost is nearly an order of magnitude higher than that of the neural network. Conversely, simpler models such as linear regression and decision trees train rapidly but exhibit significantly larger errors, indicating that they lack the expressive capacity needed to capture the nonlinear relationship between integrated dispersion and ring dimensions. Random forests and support vector machines perform moderately well but still fall short of the accuracy achieved by the neural network. Overall, the neural network offers the best balance between prediction accuracy and training time, delivering $<1$~nm precision with manageable computational effort. For this reason, we adopt neural networks as the primary modeling tool in the remainder of the manuscript.
\begin{table}[h]
\footnotesize
    \centering
    \caption{Mean width and height prediction errors and training times of six machine learning models.}    
    \begin{tabular}{cccc}
ML Model & Mean Width & Mean Height & Training \\
 &  Pred. Error (nm) & Pred. Error (nm) & time (s)\\
\hline
Linear Regression & 1.2516 &   1.7294 & 0.01 \\
Decision Trees & 8.7568  &  2.3719 & 0.05 \\
Support Vector & 3.0871  &  1.5593 & 18.7 \\
Random Forest & 5.1966  &  1.3019 & 19.3 \\
Neural Network & 0.2778 & 0.2323 & 84.1 \\
Gaussian Process & 0.0663  &  0.0526 & 277 \\
\end{tabular}
    \label{tab:ML_models}
\end{table}

In practical scenarios, experimental measurements are inevitably subject to uncertainties introduced by the finite precision of the measurement system. To emulate such conditions, we introduce controlled noise into both the training and testing datasets. Specifically, the integrated dispersion values are perturbed according to the following formula:
\begin{equation}
D_{\text{int}}^{\text{noisy}} = D_{\text{int}} + 2\times
(\text{rand}(m,\mu)-0.5) \times \Delta f,
\label{Dint_noise}
\end{equation}
where \(\text{rand}(m,\mu)\) generates uniformly distributed random values in the interval \([0,1]\), and \(\Delta f\) denotes the noise level and $m$ is the number of designs. To represent a measurement system with high precision, we set \(\Delta f = 50 \, \text{MHz}\) (consistent with the performance of many commercial laser wavemeters). Conversely, to emulate a lower-precision measurement system, we set \(\Delta f = 200 \, \text{MHz}\). These choices allow us to assess the robustness of the regression network under different levels of measurement uncertainty.

Table \ref{tab:width_errors} reports the ring width prediction errors across different gas ratios and noise conditions. In the absence of noise, the neural network achieves $<1$~nm mean absolute errors (MAE) for all gas ratios, with maximum errors of approximately 1 nm, and at most one test case exceeding a 10 nm deviation, demonstrating excellent predictive accuracy. At moderate noise levels ($\pm$50 MHz), the average error increases slightly, to a level between 0.7~nm and 2.1~nm depending on the gas ratio. Still, the coefficient of determination ($R^2$) remains above 0.99 for all gas ratios, indicating that the model can tolerate realistic perturbations in the data. For all gas ratios, there are a few cases (<3) with errors higher than 10 nm. For the $\pm$200 MHz random noise case, larger errors are observed, with MAEs reaching up to ~8.5 nm and several test cases exceeding 10 nm, especially for high gas ratios. Nevertheless, $R^2$ values remain above 0.95, confirming that the model retains high predictive fidelity under stronger noise. Finally, we note that the MxAEs (the maximum absolute errors of all the test cases) are much larger for the $\pm$200 MHz random noise case, particularly for the highest gas ratios.
\begin{table}[!h]
    \centering
    \footnotesize
    \caption{Width prediction errors for the neural network ML model. MAE: mean absolute error in nanometers. MxAE: maximum absolute error in nanometers of all the test cases. $\#\{\Delta>10 \,\text{nm}\}$: number of test cases where the error is larger than 10 nm.} 
    \begin{tabular}{c|c|c|c|c|c}    
    Gas & Noise & MAE & MxAE & R$^2$ & $\#\{\Delta>$  \\
    Ratio & Level & (nm) & (nm) & \ & $10 \ \text{nm}\}$ \\
    \hline
    \multirow{3}{*}{3:1} & No Noise & 0.278 & 1.133 & 0.99996 & 1 \\
& 50 MHz & 1.336   &   35.469  &   0.9917 & 1 \\
& 200 MHz & 2.712  &  17.027  & 0.9947 & 1 \\
\hline
\multirow{3}{*}{5:1} & No Noise & 0.172 & 0.824 & 0.99998 & 0 \\
& 50 MHz & 0.759 & 10.669  & 0.9989 & 1 \\
& 200 MHz & 5.299  & 18.052 &  0.9850 & 6 \\
\hline
\multirow{3}{*}{7:1} & No Noise & 0.259 & 1.049 & 0.99996 & 1 \\
& 50 MHz & 2.104 & 16.20 &  0.995 & 2 \\
& 200 MHz & 6.885 &  25.560  & 0.9744 & 12 \\
\hline
\multirow{3}{*}{15:1} & No Noise & 0.264 & 1.1079 & 0.99996 & 1 \\
& 50 MHz & 1.592  & 9.9694 & 0.9979 & 0 \\
& 200 MHz & 8.507  &  45.525 & 0.9504 & 18 \\
    \end{tabular}
    \label{tab:width_errors}
\end{table}

Table \ref{tab:height_errors} summarizes the ring height prediction errors. Without noise, MAEs remain well below 0.3 nm, with maximum errors near or below 1.5 nm and almost no cases exceeding 1 nm. Under $\pm$50 MHz random noise, errors increase moderately (MAEs between 0.6 nm and 1.4 nm), but $R^2$ values stay above 0.986, confirming robust performance. For the $\pm$200 MHz random noise case, deviations become more pronounced (MAEs up to ~3.4 nm and maximum errors around 20 nm), with a few test cases exceeding 10 nm. Nevertheless, $R^2$ values remain above 0.96, indicating that the model continues to provide reliable predictions even in the presence of substantial noise. Overall, height prediction errors are consistently smaller than width prediction errors; this behavior can be understood from the fact that, within the geometries studied, $D_{\rm{int}}$ is more sensitive to variations in height than in width. Consequently, the inverse mapping from dispersion to ring height is better conditioned, allowing the network to achieve higher accuracy for height prediction.

\begin{table}[!h]
    \caption{Height prediction errors for the neural network ML model.} 
    \centering
    \footnotesize
    \begin{tabular}{c|c|c|c|c|c}
    Gas & Noise & MAE & MxAE & R$^2$ & $\#\{\Delta>$  \\
    Ratio & Level & (nm) & (nm) & \ & $10 \ \text{nm}\}$ \\
    \hline
\multirow{3}{*}{3:1} & No Noise & 0.232 & 1.0018 & 0.99988 & 1 \\
& 50 MHz & 0.614 & 13.073  &  0.9957 & 1 \\
& 200 MHz & 1.037  & 10.111 &  0.9953 & 1 \\
\hline
\multirow{3}{*}{5:1} & No Noise & 0.157 & 1.4685 & 0.99992 & 1 \\
& 50 MHz & 0.587 &  7.348 &  0.9974 & 0 \\
& 200 MHz & 2.535 & 15.471 &  0.9851 & 1 \\
\hline
\multirow{3}{*}{7:1} & No Noise & 0.264 & 0.8533 & 0.99987 & 0 \\
& 50 MHz & 1.419  & 16.105 & 0.9866 & 2 \\
& 200 MHz & 3.171 & 15.397 & 0.9757 & 3 \\
\hline
\multirow{3}{*}{15:1} & No Noise & 0.274 & 0.7385 & 0.99985 & 0 \\
& 50 MHz & 1.351 & 6.760 & 0.9957 & 0 \\
& 200 MHz & 3.399 &  20.465 & 0.9647 & 4 \\
    \end{tabular}
    \label{tab:height_errors}
\end{table}

Note that the noise was added to the integrated dispersion data using a uniform (random) distribution rather than a Gaussian distribution. This choice represents a conservative and adversarial noise model, i.e., the uniform noise assigns equal probability to all perturbation amplitudes within a fixed interval, thereby increasing the likelihood of large deviations compared to Gaussian noise with the same variance. If Gaussian noise were used instead, the resulting prediction errors in the dimension inference framework would be systematically smaller. This is because Gaussian noise is strongly concentrated around its mean (zero), with large-amplitude perturbations occurring with exponentially decreasing probability. As a result, most dispersion samples would experience only small perturbations, preserving the smooth spectral structure and curvature of the integrated dispersion profile that are critical for accurate dimension extraction. In contrast, uniform noise introduces larger amplitude fluctuations that are uncorrelated across modes, leading to stronger distortion of higher-order dispersion features. These features play a dominant role in constraining geometric parameters, and their degradation directly propagates into larger prediction errors.

To address the potential impact of experimental frequency shifts and fabrication-induced deviations on the ring dimension prediction framework, we conducted an additional sensitivity analysis by introducing stochastic perturbations to the resonance frequencies. Recognizing that experimental jitter and thermal fluctuations can shift resonance positions, we simulated two distinct noise regimes by adding random frequency deviations of $\pm 50$ MHz and $\pm 200$ MHz to the calculated resonance frequencies in our test dataset. To assess the architecture's robustness, we first trained a neural network (a three-layer deep regression neural network utilizing 256 neurons per hidden layer, Leaky ReLU activation functions, and a 20~\% dropout rate for regularization, optimized via the Adam algorithm over 200 epochs) to predict the free spectral range, $D_1$, from the integrated dispersion dataset using our noise-free training set (SM$_2$). We then tested it on our noise-added datasets. At a noise floor of 50 MHz, the model achieved an average root mean squared prediction error for $D_1$ of 0.079\%. When we used our dimension-prediction neural network on the integrated dispersion test dataset with slightly off $D_1$ values, we observed average errors of 0.18 nm for the ring width and 0.12 nm for the ring height. Our results demonstrate that the neural network exhibits high resilience to frequency noise, functioning as a nonlinear filter that prioritizes the global curvature of the dispersion profile over localized stochastic fluctuations. Remarkably, quadrupling the noise level to 200 MHz only marginally increased the $D_1$ prediction error to 0.088\%, while the width and height average errors were determined to be 0.29 nm and 0.23 nm, respectively.  The observed stability in $D_1$ prediction, despite a fourfold increase in frequency noise, can be attributed to the scale disparity between the optical frequencies and the MHz-scale perturbations. Before proceeding to our next study, it is further worth noting that a neural network with only three hidden layers can predict $D_1$ from ring width and height with remarkable precision. Even with $\pm 50$ MHz of added noise, the average prediction error remains as low as 0.0139~\%. When the noise level is quadrupled to 200 MHz, the error rate increases only marginally to 0.0222~\%, demonstrating the model's high robustness.

\begin{figure}[!t]
    \centering
    \includegraphics[width=0.5\linewidth]{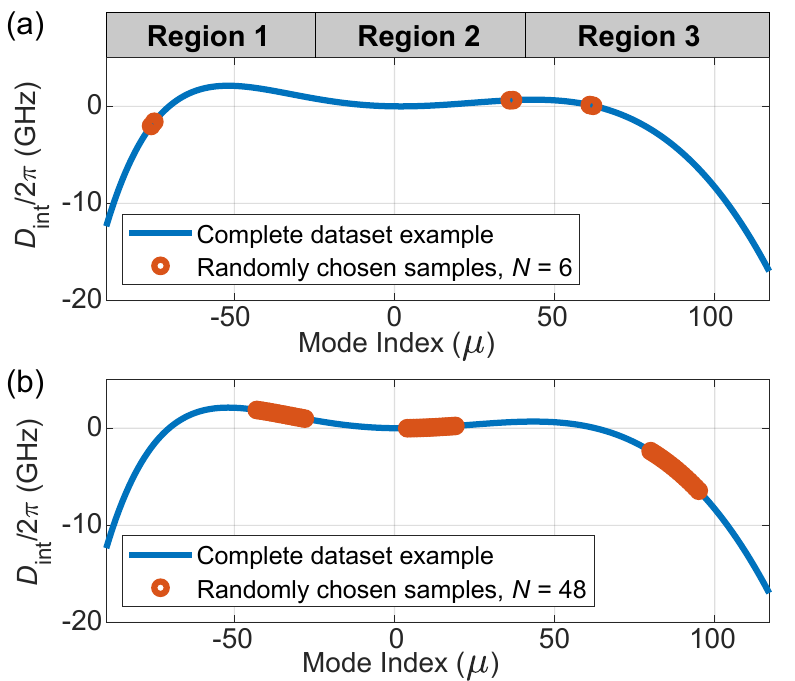}
    \caption{Examples of random sampling from the integrated dispersion dataset, where $M$ random samples are chosen from each of regions 1 ($\mu=-91$ to $\mu=-25$), 2 ($\mu=-24$ to $\mu=41$), and 3 ($\mu=42$ to $\mu=109$), for a total of $N=3M$ samples. 
    (a) Case for $M = 2$ ($N = 6$), and (b) case for $M = 16$ ($N = 48$). 
    The blue line represents the complete $D_{\text{int}}/2\pi$ dataset, while the red circles denote $M$ neighboring samples randomly chosen in three regions and used for training and testing.}
    \label{fig:Dint_sampling}
\end{figure}

For the initial study, we utilized the complete integrated dispersion dataset. To further investigate how many $D_{\text{int}}$ samples are required for accurate predictions, we designed a statistical sampling procedure. Specifically, we randomly select $M$ number of $D_{\text{int}}$ samples from each of three regions: region~1, region~2, and region~3, which include 67 modes from the left (mode indices from $-91$ to $-25$), middle (from $-24$ to $41$), and right (from $42$ to $109$) portions of the spectrum, respectively (Fig.~\ref{fig:Dint_sampling}). This results in a total of $N = 3M$ sampled data points. The random selection process was repeated 40 times to emulate a Monte Carlo–type analysis, allowing us to assess statistical variability. The parameter $M$ was systematically varied from 2 to 16 in increments of 2, corresponding to total sample sizes $N$ ranging from 6 to 48 in steps of 6. Two representative examples representing the lower and upper limits of this sampling are illustrated in Fig.~\ref{fig:Dint_sampling}: (a) $M = 2$ ($N = 6$) and (b) $M = 16$ ($N = 48$).

\begin{figure}[!t]
    \centering
    \includegraphics[width=0.5\linewidth]{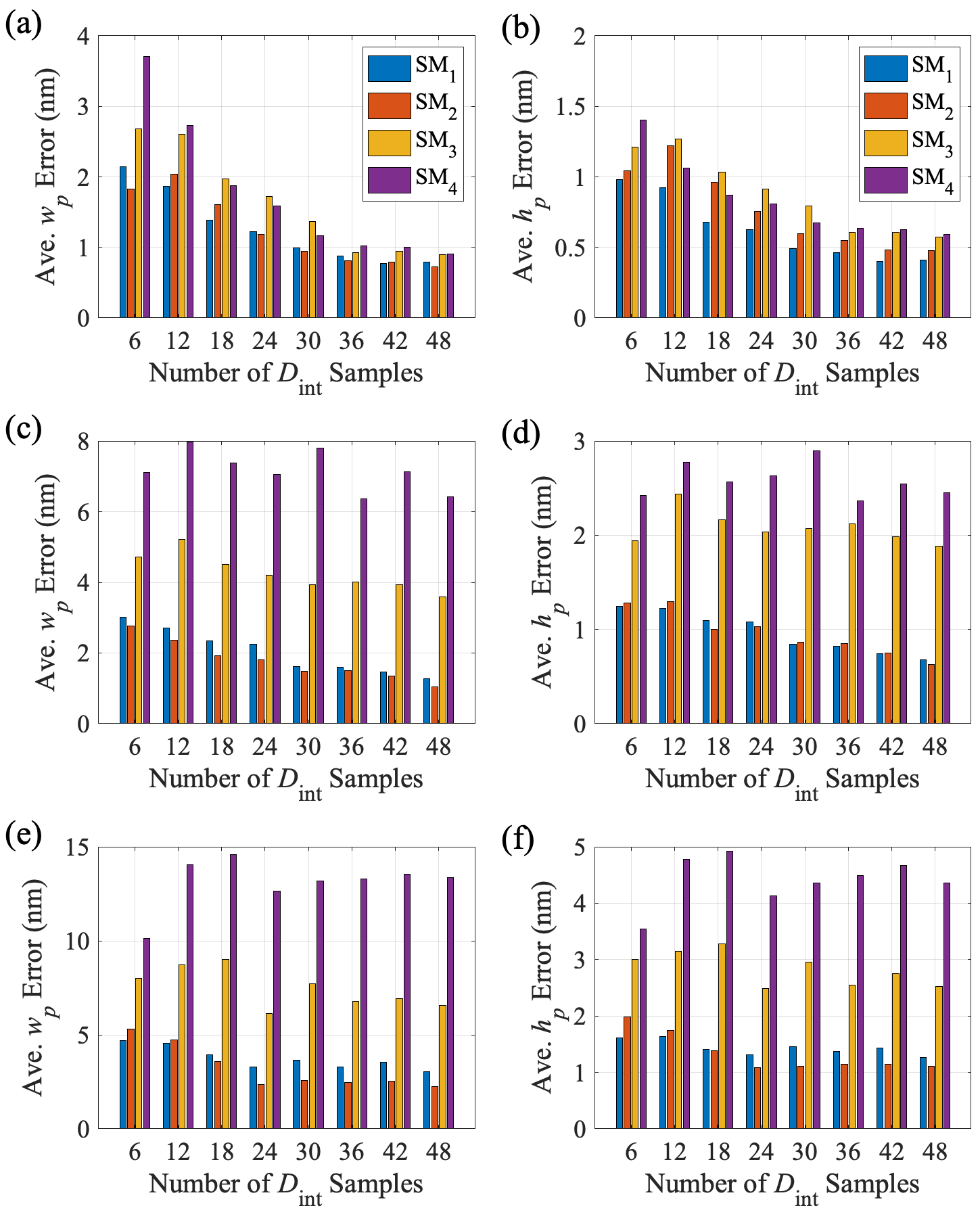}
    \caption{Left (right) column: average width (height) prediction error as a function of number of $D_{\rm{int}}$ samples used in the neural network training and testing. The top, middle, and bottom rows are the cases with no noise, $\pm$50 MHz random noise, and $\pm$200 MHz random noise, respectively.}
    \label{fig:N_vs_regression_error}
\end{figure}

The results in Fig.~\ref{fig:N_vs_regression_error} demonstrate that prediction accuracy strongly depends on the number of $D_{\text{int}}$ samples, noise level, and the gas ratio. Left column figures, panels (a), (c), and (e), report the average width prediction error, while (b), (d), and (f) in the right column show the height prediction error, with the rows corresponding to no noise, $\pm 50$~MHz noise, and $\pm 200$~MHz noise, respectively. 

When there is no noise, using only six $D_\text{int}$ samples produces ring width and height errors that in all cases are $<$4~nm and $<1.5$~nm, respectively, with the errors reaching
$<1$~nm with $>40$ $D_{\text{int}}$ samples. In all cases, a larger $N$ improves accuracy, confirming that broader sampling reduces statistical variability. Once noise is added, the above trends hold for the SM$_1$ and SM$_2$ gas ratio cases, with the ring width and height errors ranging from $<$5~nm and $<2$~nm ($N$=6), respectively, to $<3$~nm and $<1.5$~nm ($N>40$), respectively. However, for the noisy cases, we do not observe the same effect for the moderate and high gas ratios (SM$_3$ and SM$_4$). As mentioned earlier, due to the flattening of the $D_{\text{int}}$ spectrum with increasing gas ratios, the prediction errors also increase with increasing gas ratios.

Going forward, the case $N=45$, i.e., 15 $D_{\text{int}}$ samples recorded at wavelengths closer to the pump wavelength and two other ends of the spectrum, is especially important since it resembles our experimental setting (discussed later). Under realistic noisy conditions, $N=45$ yields width errors in the 2~nm to 8~nm range and height errors in the 1~nm to 3~nm for $\pm 50$~MHz random noise. For $\pm 200$~MHz random noise cases, the average width and height prediction errors are in the 4~nm to 14~nm and 1~nm to 5~nm, respectively. Overall, we can conclude that for the more precise measurement systems, 20 or more $D_{\text{int}}$ is sufficient to predict the ring dimensions with errors less than 8 nm, and having more $D_{\text{int}}$ samples helps with decreasing the prediction errors, especially if the NH$_3$-to-SiH$_4$ gas-ratio kept low. For less precise measurement systems, one should expect higher errors. Moreover, as noted above, the performance with a significantly reduced number of samples (e.g., $N=6$) is still good for the SM$_1$ and SM$_2$ cases, with increased measurement throughput of potential practical interest. 

\begin{figure}[!h]
     \centering
     \includegraphics[width=0.5\linewidth]{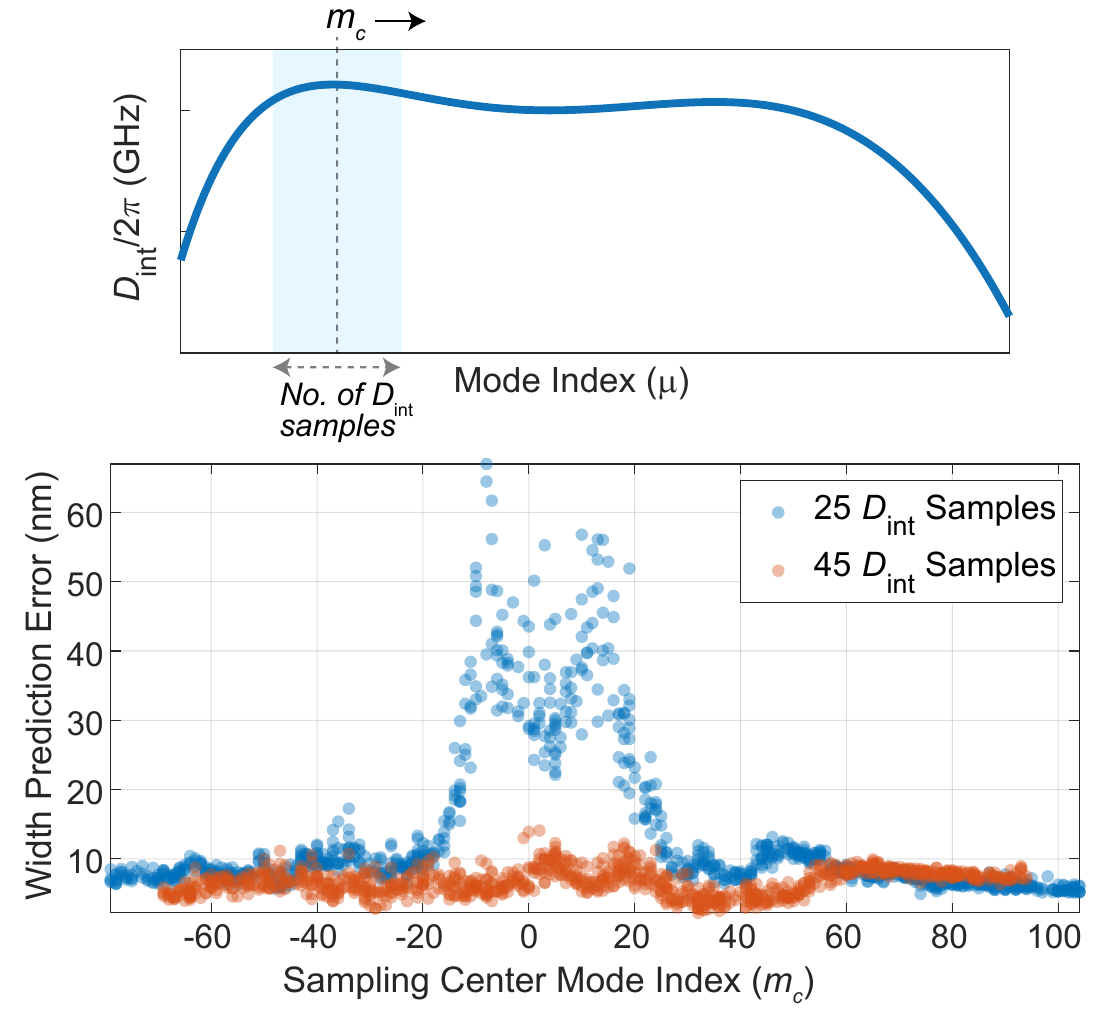}
     \caption{Width prediction error as a function of the sampling center mode number $m_c$, obtained from 1000 trials using the Sellmeier model-1 dataset with 50~MHz random noise. For each trial, 25 (blue) and 45 (red)  consecutive \(D_{\mathrm{int}}\) values centered at \(m_c\) were used to train and test the neural network model. See the top schematic for definitions of $m_c$ and the number of $D_\text{int}$ samples. The resonator $D_1$ value is assumed to be well-understood independent of where the sampling takes place.}
     \label{fig:mc_vs_accuracy}
 \end{figure}

We also investigate the impact of the spectral sampling location on the accuracy of width predictions using synthetic data generated from the SM$_2$ dataset with added 50~MHz random noise. To probe this effect, we constructed sub-datasets by selecting either 25 or 45 consecutive $D_{\mathrm{int}}$ values centered around a mode index $m_c$, repeating the process 1000 times for each $m_c$ to account for variability due to neural network initialization. Note that 25 and 45 consecutive $D_{\mathrm{int}}$ values correspond to $\approx 10~\%$ and $\approx 20~\%$ of the entire octave-spanning spectrum (750~nm to 1600~nm), respectively. The results in Fig.~\ref{fig:mc_vs_accuracy} reveal a strong dependence on sampling location when only 25 samples are used. In particular, if the sampling window is centered near the pump frequency, where $D_{\mathrm{int}}$ values are small and nearly flat, the network receives little discriminative information and prediction errors rise sharply. By contrast, when the window is shifted away from the pump into regions where $D_{\mathrm{int}}$ exhibits stronger curvature, the errors drop substantially, typically to the 8~nm to 15~nm range. Crucially, when the sampling window covers $\approx 20~\%$ of the spectrum (45 samples), every window inevitably includes some portion of sloped $D_{\mathrm{int}}$. This demonstrates that reliable predictions require not only a sufficient number of samples, but also that at least some of these samples be chosen away from the pump frequency, where the spectral features are more informative. That being said, we note that an implicit assumption of the above analysis is that the resonator $D_1$ coefficient (i.e., the free-spectral range around the pump) is accurately known throughout.

\subsection{Sellmeier Model Classification}
The second network, Fig.~\ref{fig:networks}(b), handles a classification problem in which the goal is to identify the Sellmeier model used for the Si$_3$N$_4$'s relative electrical permittivity (refractive index) in $D_\text{int}$ calculations. In this case, the same normalized $D_\text{int}$ vectors serve as input, while the output is categorical, corresponding to one of four possible Sellmeier models (Table \ref{tab:si3n4_disp}). The architecture comprises four hidden layers with progressively decreasing sizes of 128, 64, 32, and 16 neurons, respectively. The first two layers employed hyperbolic tangent (tanh) activations, while the subsequent two layers utilized rectified linear unit (ReLU) activations. A softmax layer with four neurons was used at the output to generate class probabilities. Training was conducted using the Adam optimizer with a categorical cross-entropy loss function, a batch size of 64, and 100 epochs.

\begin{figure}[!h]
    \centering
    \includegraphics[width=0.4\linewidth]{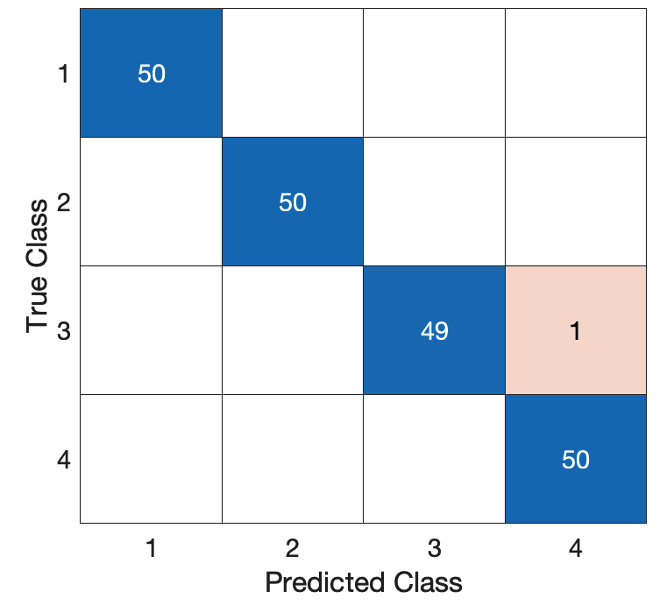}
    \caption{Confusion matrix for the Sellmeier model classification task under all cases (no noise, 50~MHz noise, and 200~MHz noise).}
    \label{fig:confusion_charts}
\end{figure}

For the initial study, the training and testing inputs are $1764\times201$ and $200\times201$, respectively; the training and testing outputs are $1764\times1$ and $200\times1$, respectively. Figure~\ref{fig:confusion_charts} presents the confusion matrix that was obtained under all varying noise conditions. In all cases, the classifier achieves nearly perfect accuracy, with only a single misclassification observed between classes 3 and 4 among 200 test cases. This result shows that, unlike the regression problem, where we predict the ring dimensions, the classification task of identifying the Sellmeier model type is considerably more robust to noise. This robustness arises because different gas ratios (as described by Sellmeier models) generate markedly distinct $D_{\mathrm{int}}$ profiles.

\begin{figure}[!h]
    \centering
    \includegraphics[width=0.4\linewidth]{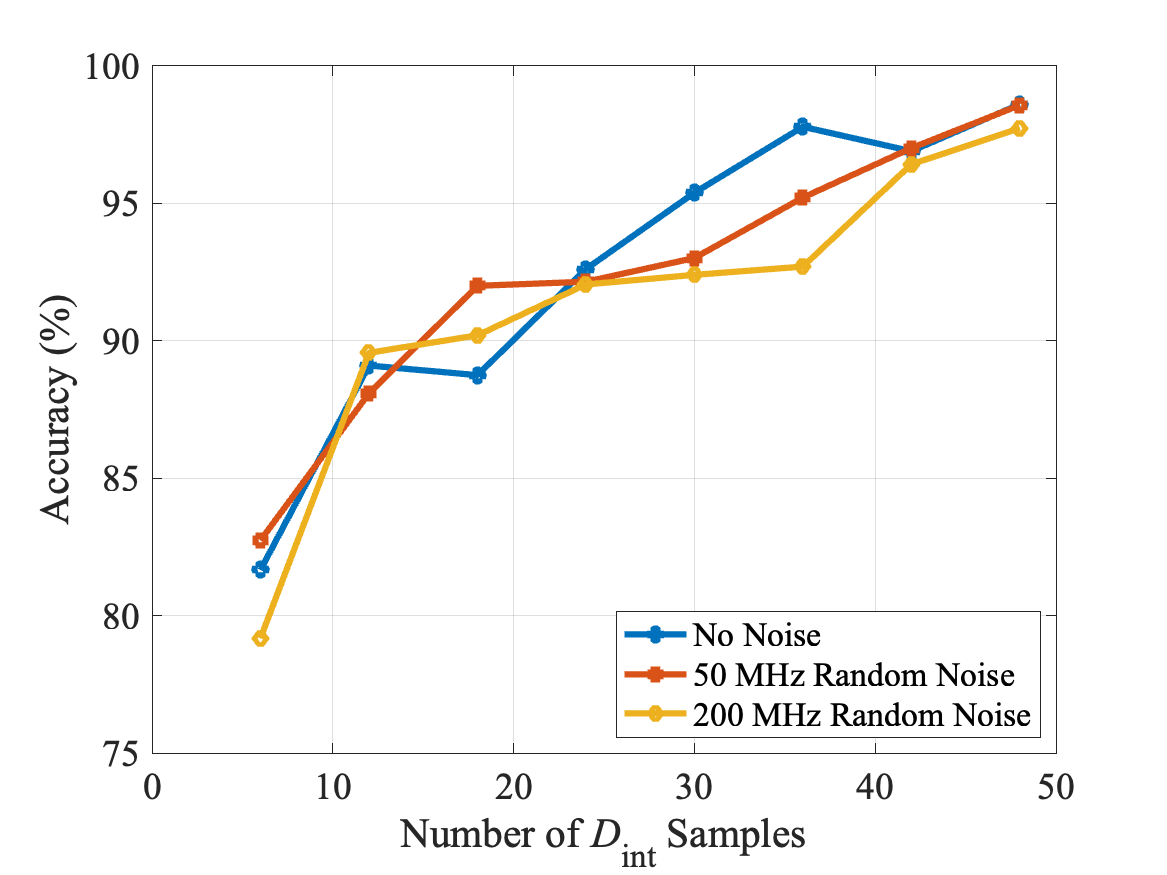}
    \caption{Average Sellmeier model classification accuracy as a function of the total number of $D_{\text{int}}$ samples, $N$, under no noise, $\pm 50$~MHz random noise, and $\pm 200$~MHz random noise.}
    \label{fig:mean_classification_acc_vs_N}
\end{figure}

To determine the minimum number of $D_{\text{int}}$ samples required for accurate classification, we follow a similar strategy as in the previous section: we randomly choose $M$ samples in three regions, yielding a total of $N = 3M$ samples. Figure~\ref{fig:mean_classification_acc_vs_N} shows the mean classification accuracy as a function of $N$ under three noise conditions: no noise, 50~MHz random noise, and 200~MHz random noise. For small sample sizes ($N \leq 12$), the accuracy varies widely, ranging from 80~\% to 90~\%, reflecting the limited discriminative information provided by only a few dispersion points. As $N$ increases, the accuracy improves consistently for all cases. With approximately $N = 21$ samples, classification accuracy already exceeds 90~\%, even in the presence of 200 MHz noise. Beyond this point, the curves rise gradually, with all three noise conditions converging to above 97~\% accuracy once $N \gtrsim 45$. Importantly, the no-noise and 50~MHz noise cases track each other closely across all $N$, while the 200~MHz noise case shows slightly lower performance for small $N$ but catches up as the number of samples increases. These results highlight the robustness of the classifier against noise and demonstrate that accurate classification can be achieved with a relatively modest number of $D_{\text{int}}$ measurements ($N \approx 21$ to 30), without requiring the complete spectral information.

\subsection{$D_{\rm{int}}$ Spectrum Prediction}
In our final study, we employ the Sellmeier model for Si$_3$N$_4$ from \cite{Luke:15} and generate another synthetic dataset following the procedure described in Section~\ref{DintCalculation}. The outer ring radius is fixed at $23~\mu$m, while the ring height and width are varied to construct a dataset for integrated dispersion spectrum prediction. Specifically, the height is swept from $620$~nm to $720$~nm in increments of $10$~nm, and the width is swept from $800$~nm to $900$~nm in increments of $10$~nm, resulting in $121$ unique training designs.

After building our training dataset, we construct a neural network model, shown in Fig. \ref{fig:networks} (c), to predict the integrated dispersion spectrum of a microring resonator directly from ring width and height. To achieve this, we first normalized the simulated dispersion data and then approximated each spectrum using a sixth-order polynomial, thereby reducing the high-dimensional spectral response to a compact representation defined by seven polynomial coefficients. These coefficients, once normalized to ensure numerical stability, served as the output training targets, while the corresponding ring dimensions provided the input features. The neural network architecture was designed with three fully connected layers (64, 128, and 128 nodes), leaky rectified linear unit activations, batch normalization, and moderate dropout rates (0.1, 0.1, and 0.2) to balance model complexity with generalization ability. The training procedure employed the Adam optimizer with an adaptive learning rate schedule, early stopping criteria, and $L_2$ regularization to mitigate overfitting. Following training, the network was used to infer polynomial coefficients for given test dimensions, and the resulting coefficients were evaluated with a polynomial expansion to reconstruct the full dispersion spectrum. 

\begin{figure}[h!]
    \centering
    \includegraphics[width=0.6\linewidth]{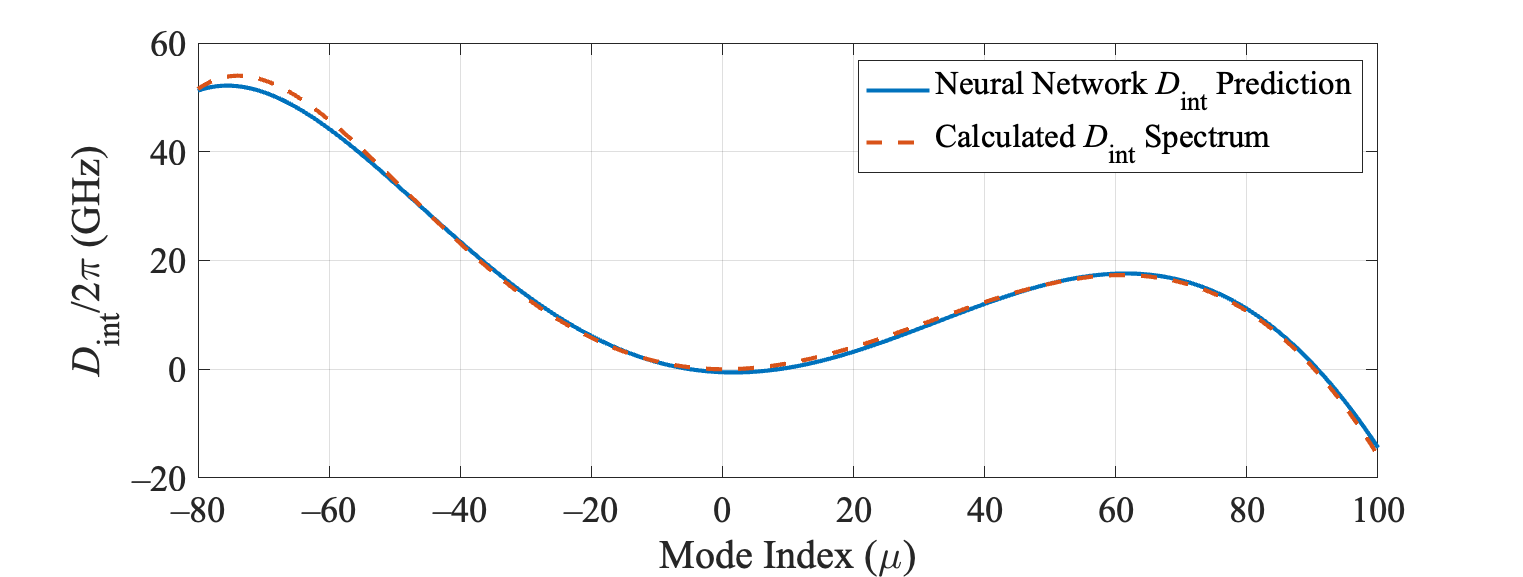}
    \caption{Comparison of dispersion spectra obtained from neural network prediction (blue) and numerical simulation (red dashed curve) for randomly chosen $RW$ and $RH$ values of $867.9~\text{nm}$ and $664.2~\text{nm}$, respectively.}
    \label{Dint_prediction_from_dimensions}
\end{figure}

The red dashed curve in Fig. \ref{Dint_prediction_from_dimensions} shows the integrated dispersion spectrum  calculated for randomly chosen dimensions of $RW = 867.9~\text{nm}$ and $RH = 664.2~\text{nm}$. When the spectrum prediction network was queried with the same dimensions, we obtained the spectrum shown with a blue solid curve in the same figure. This close agreement and many other successful test cases which are not shown here indicate that the network successfully captured the nonlinear mapping between ring geometry and dispersion profile without introducing spurious oscillations, thereby providing a practical tool for inverse design and accurate dispersion engineering of integrated photonic resonators.

\section{Conclusion}
In this work, we demonstrate a non-destructive method for dispersion metrology of foundry-fabricated devices based on the optical characterization of the integrated dispersion ($D_\mathrm{int}(\mu)$) of microring resonators. Importantly, we show that through proper training of a machine-learning model, a limited sample of judiciously chosen resonance frequencies can enable accurate dimension prediction even in the presence of realistic measurement noise.
First, we demonstrated that $D_\mathrm{int}(\mu)$ profiles provide a compact and information-rich fingerprint for both regression and classification tasks in Si$_3$N$_4$ microring resonators. Using numerically generated datasets, we trained neural networks to predict device dimensions and fabrication conditions with high accuracy. For the regression task, the networks achieved sub-nanometer mean absolute errors in the absence of noise and maintained reliable performance under realistic perturbations, with width predictions being more sensitive to noise than height predictions. For the classification task, identifying the Sellmeier model type associated with precursor gas ratios, accuracies exceeding 99~\% were obtained and shown to be robust to added noise, reflecting the distinctiveness of the $D_{\mathrm{int}}$ profiles across different gas ratios. Importantly, we found that dispersion samples taken far from the pump frequency are most informative: with a precise measurement system of $\pm 50$~MHz deviation, approximately 45 samples are sufficient to predict device dimensions with errors below 8~nm, while for less precise systems with $\pm 200$~MHz deviation, the expected errors approximately double. %
Having demonstrated such an approach, we reversed it and also demonstrated the prediction of the $D_{\mathrm{int}}$ profiles directly from ring dimensions, which provides a complementary capability for device modeling and design. Altogether, these findings establish integrated dispersion as a powerful observable for non-destructive and rapid quality control in large-scale photonic manufacturing, and they offer a generalizable framework for linking measurable optical responses to underlying fabrication parameters, thereby accelerating the development and deployment of advanced photonic systems.

\section*{Data Availability}
The codes and datasets generated during and analyzed during the current study are available in the GitHub repository, \url{https://github.com/simsekergun/DeviceMetrology}.

\bibliographystyle{ieeetr}
\bibliography{references}

@article{fahrenkopf2019aim,
  title={The {AIM} photonics {MPW}: A highly accessible cutting edge technology for rapid prototyping of photonic integrated circuits},
  author={Fahrenkopf, Nicholas M and McDonough, Colin and Leake, Gerald L and Su, Zhan and Timurdogan, Erman and Coolbaugh, Douglas D},
  journal={IEEE Journal of Selected Topics in Quantum Electronics},
  volume={25},
  number={5},
  pages={1--6},
  year={2019},
  publisher={IEEE}
}

@article{Die_Ring_Solver,
author = {Simsek, Ergun and Niang, Alioune and Islam, Raonaqul and Courtright, Logan and Shandilya, Pradyoth and Carter, Gary M. and Menyuk, Curtis R.},
date = {2025/10/08},
doi = {10.1038/s41598-025-18869-z},
id = {Simsek2025},
isbn = {2045-2322},
journal = {Scientific Reports},
number = {1},
pages = {35098},
title = {A mixed-field formulation for modeling dielectric ring resonators and its application in optical frequency comb generation},
url = {https://doi.org/10.1038/s41598-025-18869-z},
volume = {15},
year = {2025}}

@article{Moille:21,
author = {Gregory Moille and Daron Westly and Gregory Simelgor and Kartik Srinivasan},
journal = {Opt. Lett.},
number = {23},
pages = {5970--5973},
publisher = {Optica Publishing Group},
title = {Impact of the precursor gas ratio on dispersion engineering of broadband silicon nitride microresonator frequency combs},
volume = {46},
month = {Dec},
year = {2021},
url = {https://opg.optica.org/ol/abstract.cfm?URI=ol-46-23-5970},
doi = {10.1364/OL.440907},
}

@article{TAN1998158,
title = {Determination of refractive index of silica glass for infrared wavelengths by {IR} spectroscopy},
journal = {Journal of Non-Crystalline Solids},
volume = {223},
number = {1},
pages = {158-163},
year = {1998},
issn = {0022-3093},
doi = {https://doi.org/10.1016/S0022-3093(97)00438-9},
url = {https://www.sciencedirect.com/science/article/pii/S0022309397004389},
author = {C.Z. Tan}
}

@ARTICLE{LMoptimization,
  author={Wilamowski, Bogdan M. and Yu, Hao},
  journal={IEEE Transactions on Neural Networks}, 
  title={Improved Computation for {Levenberg–Marquardt} Training}, 
  year={2010},
  volume={21},
  number={6},
  pages={930-937},
  doi={10.1109/TNN.2010.2045657}}

@article{Ahn2022,
	author = {Ahn, Geun Ho and Yang, Ki Youl and Trivedi, Rahul and White, Alexander D. and Su, Logan and Skarda, Jinhie and Vu{\v c}kovi{\'c}, Jelena},
	date = {2022/06/15},
	doi = {10.1021/acsphotonics.2c00020},
	journal = {ACS Photonics},
	journal1 = {ACS Photonics},
	journal2 = {ACS Photonics},
	month = {06},
	n2 = {The automation of device design enabled by optimization and machine learning techniques has been transformative for photonics. While this automation has been successful for nonresonant devices, automated photonic design has remained elusive for resonant devices, key elements for on-chip communication technologies of biosensing and quantum optics, due to their highly nonconvex optimization landscapes. We propose a framework that solves this problem by mapping the design of photonic resonators to a set of nonresonant design problems. We theoretically and experimentally demonstrate this framework and show flexible dispersion engineering, a quality factor beyond 2 million on silicon-on-insulator with single-mode operation, and selective wavelength-band operation.},
	number = {6},
	pages = {1875--1881},
	publisher = {American Chemical Society},
	title = {Photonic Inverse Design of On-Chip Microresonators},
	type = {doi: 10.1021/acsphotonics.2c00020},
	url = {https://doi.org/10.1021/acsphotonics.2c00020},
	volume = {9},
	year = {2022},
	year1 = {2022}}

@Article{Wang2021,
AUTHOR = {Wang, Zhaonian and Du, Jiangbing and Shen, Weihong and Liu, Jiacheng and He, Zuyuan},
TITLE = {Efficient Design for Integrated Photonic Waveguides with Agile Dispersion},
JOURNAL = {Sensors},
VOLUME = {21},
YEAR = {2021},
NUMBER = {19},
ARTICLE-NUMBER = {6651},
URL = {https://www.mdpi.com/1424-8220/21/19/6651},
PubMedID = {34640972},
ISSN = {1424-8220},
DOI = {10.3390/s21196651}
}

@article{Pal:23,
author = {Arghadeep Pal and Alekhya Ghosh and Shuangyou Zhang and Toby Bi and Pascal Del'Haye},
journal = {Opt. Express},
number = {5},
pages = {8020--8028},
publisher = {Optica Publishing Group},
title = {Machine learning assisted inverse design of microresonators},
volume = {31},
month = {Feb},
year = {2023},
url = {https://opg.optica.org/oe/abstract.cfm?URI=oe-31-5-8020},
doi = {10.1364/OE.479899},
}

@article{DelHaye:2007,
  author = {P. Del'Haye and A. Schliesser and O. Arcizet and T. Wilken and R. Holzwarth and T. J. Kippenberg},
  title = {Optical frequency comb generation from a monolithic microresonator},
  journal = {Nature},
  volume = {450},
  pages = {1214--1217},
  year = {2007},
  doi = {10.1038/nature06401}
}

@article{Kippenberg:2018,
  author = {T. J. Kippenberg and A. L. Gaeta and M. Lipson and M. L. Gorodetsky},
  title = {Dissipative Kerr solitons in optical microresonators},
  journal = {Science},
  volume = {361},
  number = {6402},
  year = {2018},
  doi = {10.1126/science.aan8083}
}

@article{Moss:2013,
  author = {D. J. Moss and R. Morandotti and A. L. Gaeta and M. Lipson},
  title = {New {CMOS}-compatible platforms based on silicon nitride and {Hydex} for nonlinear optics},
  journal = {Nat. Photonics},
  volume = {7},
  pages = {597--607},
  year = {2013},
  doi = {10.1038/nphoton.2013.183}
}

@inproceedings{simsek2025stop,
  title={One-Stop-Shop for Modeling Optical Frequency Comb Generation},
  author={Simsek, E. and Niang, A. and Shandilya, R. P. and Islam and Courtright, L. and Carter, G. M. and Menyuk, C. R.},
  booktitle={26th International Conference on Electromagnetics in Advanced Applications (ICEAA 2025)},
  address={Palermo, Italy},
  month={September},
  year={2025},
  note={The conference proceedings will be available in late 2025}
}

@article{ucombs_for_quantum,
  title={Quantum optical microcombs},
  author={Kues, Michael and Reimer, Christian and Lukens, Joseph M and Munro, William J and Weiner, Andrew M and Moss, David J and Morandotti, Roberto},
  journal={Nature Photonics},
  volume={13},
  number={3},
  pages={170--179},
  year={2019},
  publisher={Nature Publishing Group UK London},
  url = {https://doi.org/10.1038/s41566-019-0363-0}
}

@article{ucombs_for_comm,
  title={Coherent terabit communications with microresonator Kerr frequency combs},
  author={Pfeifle, Joerg and Brasch, Victor and Lauermann, Matthias and Yu, Yimin and Wegner, Daniel and Herr, Tobias and Hartinger, Klaus and Schindler, Philipp and Li, Jingshi and Hillerkuss, David and others},
  journal={Nature photonics},
  volume={8},
  number={5},
  pages={375--380},
  year={2014},
  publisher={Nature Publishing Group UK London},
  url = {https://doi.org/10.1038/nphoton.2014.57}
}

@article{hansch2,
	author = {Udem, Th. and Holzwarth, R. and H{\"a}nsch, T. W.},
	date = {2002/03/01},
	doi = {10.1038/416233a},
	id = {Udem2002},
	isbn = {1476-4687},
	journal = {Nature},
	number = {6877},
	pages = {233--237},
	title = {Optical frequency metrology},
	url = {https://doi.org/10.1038/416233a},
	volume = {416},
	year = {2002}}

@article{ES_Attention:2023,
	author = {Soroush, Masoud and Simsek, Ergun and Moille, Gregory and Srinivasan, Kartik and Menyuk, Curtis R.},
	date = {2023/06/21},
	doi = {10.1021/acsphotonics.3c00054},
	journal = {ACS Photonics},
	month = {06},
	number = {6},
	pages = {1795--1805},
	publisher = {American Chemical Society},
	title = {Predicting Broadband Resonator-Waveguide Coupling for Microresonator Frequency Combs through Fully Connected and Recurrent Neural Networks and Attention Mechanism},
	type = {doi: 10.1021/acsphotonics.3c00054},
	url = {https://doi.org/10.1021/acsphotonics.3c00054},
	volume = {10},
	year = {2023},
	year1 = {2023}}

@article{SO2025,
author = {Shao-Chien Ou and Alin O. Antohe and Lewis G. Carpenter and Gregory Moille and Kartik Srinivasan},
journal = {Opt. Lett.},
number = {18},
pages = {5578--5581},
publisher = {Optica Publishing Group},
title = {300 mm wafer-scale {SiN} platform for broadband soliton microcombs compatible with alkali atomic references},
volume = {50},
month = {Sep},
year = {2025},
url = {https://opg.optica.org/ol/abstract.cfm?URI=ol-50-18-5578},
doi = {10.1364/OL.571893},
}

@article{gaeta_photonic-chip-based_2019,
	title = {Photonic-chip-based frequency combs},
	volume = {13},
	issn = {1749-4885, 1749-4893},
	url = {http://www.nature.com/articles/s41566-019-0358-x},
	doi = {10.1038/s41566-019-0358-x},
	language = {en},
	number = {3},
	urldate = {2020-03-11},
	journal = {Nature Photonics},
	author = {Gaeta, Alexander L. and Lipson, Michal and Kippenberg, Tobias J.},
	month = mar,
	year = {2019},
	pages = {158--169},
}

@article{li_stably_2017,
	title = {Stably accessing octave-spanning microresonator frequency combs in the soliton regime},
	volume = {4},
	issn = {2334-2536},
	url = {https://www.osapublishing.org/abstract.cfm?URI=optica-4-2-193},
	doi = {10.1364/OPTICA.4.000193},
	language = {en},
	number = {2},
	urldate = {2021-10-13},
	journal = {Optica},
	author = {Li, Qing and Briles, Travis C. and Westly, Daron A. and Drake, Tara E. and Stone, Jordan R. and Ilic, B. Robert and Diddams, Scott A. and Papp, Scott B. and Srinivasan, Kartik},
	month = feb,
	year = {2017},
	pages = {193},
}

@article{briles_interlocking_2018,
	title = {Interlocking {Kerr}-microresonator frequency combs for microwave to optical synthesis},
	volume = {43},
	issn = {0146-9592, 1539-4794},
	url = {https://www.osapublishing.org/abstract.cfm?URI=ol-43-12-2933},
	doi = {10.1364/OL.43.002933},
	language = {en},
	number = {12},
	urldate = {2021-10-13},
	journal = {Optics Letters},
	author = {Briles, Travis C. and Stone, Jordan R. and Drake, Tara E. and Spencer, Daryl T. and Fredrick, Connor and Li, Qing and Westly, Daron and Ilic, B. R. and Srinivasan, Kartik and Diddams, Scott A. and Papp, Scott B.},
	month = jun,
	year = {2018},
	pages = {2933},
}

@article{spencer_optical-frequency_2018,
	title = {An optical-frequency synthesizer using integrated photonics},
	volume = {557},
	issn = {0028-0836, 1476-4687},
	url = {http://www.nature.com/articles/s41586-018-0065-7},
	doi = {10.1038/s41586-018-0065-7},
	language = {en},
	number = {7703},
	urldate = {2021-10-13},
	journal = {Nature},
	author = {Spencer, Daryl T. and Drake, Tara and Briles, Travis C. and Stone, Jordan and Sinclair, Laura C. and Fredrick, Connor and Li, Qing and Westly, Daron and Ilic, B. Robert and Bluestone, Aaron and Volet, Nicolas and Komljenovic, Tin and Chang, Lin and Lee, Seung Hoon and Oh, Dong Yoon and Suh, Myoung-Gyun and Yang, Ki Youl and Pfeiffer, Martin H. P. and Kippenberg, Tobias J. and Norberg, Erik and Theogarajan, Luke and Vahala, Kerry and Newbury, Nathan R. and Srinivasan, Kartik and Bowers, John E. and Diddams, Scott A. and Papp, Scott B.},
	month = may,
	year = {2018},
	pages = {81--85},
}

@article{newman_architecture_2019,
	title = {Architecture for the photonic integration of an optical atomic clock},
	volume = {6},
	issn = {2334-2536},
	url = {https://www.osapublishing.org/abstract.cfm?URI=optica-6-5-680},
	doi = {10.1364/OPTICA.6.000680},
	language = {en},
	number = {5},
	urldate = {2021-10-13},
	journal = {Optica},
	author = {Newman, Zachary L. and Maurice, Vincent and Drake, Tara and Stone, Jordan R. and Briles, Travis C. and Spencer, Daryl T. and Fredrick, Connor and Li, Qing and Westly, Daron and Ilic, B. R. and Shen, Boqiang and Suh, Myoung-Gyun and Yang, Ki Youl and Johnson, Cort and Johnson, David M. S. and Hollberg, Leo and Vahala, Kerry J. and Srinivasan, Kartik and Diddams, Scott A. and Kitching, John and Papp, Scott B. and Hummon, Matthew T.},
	month = may,
	year = {2019},
	pages = {680},
}

@article{chang_integrated_2022,
	title = {Integrated optical frequency comb technologies},
	volume = {16},
	issn = {1749-4885, 1749-4893},
	url = {https://www.nature.com/articles/s41566-021-00945-1},
	doi = {10.1038/s41566-021-00945-1},
	language = {en},
	number = {2},
	urldate = {2022-02-07},
	journal = {Nature Photonics},
	author = {Chang, Lin and Liu, Songtao and Bowers, John E.},
	month = feb,
	year = {2022},
	pages = {95--108},
}

@article{pfeiffer_photonic_2016,
	title = {Photonic {Damascene} process for integrated high-{Q} microresonator based nonlinear photonics},
	volume = {3},
	copyright = {https://doi.org/10.1364/OA\_License\_v1\#VOR-OA},
	issn = {2334-2536},
	url = {https://opg.optica.org/abstract.cfm?URI=optica-3-1-20},
	doi = {10.1364/OPTICA.3.000020},
	language = {en},
	number = {1},
	urldate = {2025-11-22},
	journal = {Optica},
	author = {Pfeiffer, Martin H. P. and Kordts, Arne and Brasch, Victor and Zervas, Michael and Geiselmann, Michael and Jost, John D. and Kippenberg, Tobias J.},
	month = jan,
	year = {2016},
	pages = {20},
}

@article{ji_methods_2021,
	title = {Methods to achieve ultra-high quality factor silicon nitride resonators},
	volume = {6},
	issn = {2378-0967},
	url = {https://aip.scitation.org/doi/10.1063/5.0057881},
	doi = {10.1063/5.0057881},
	language = {en},
	number = {7},
	urldate = {2021-10-21},
	journal = {APL Photonics},
	author = {Ji, Xingchen and Roberts, Samantha and Corato-Zanarella, Mateus and Lipson, Michal},
	month = jul,
	year = {2021},
	pages = {071101},
}

@article{moille_all-optical_2025,
	title = {All-optical noise quenching of an integrated frequency comb},
	volume = {12},
	copyright = {https://doi.org/10.1364/OA\_License\_v2\#VOR-OA},
	issn = {2334-2536},
	url = {https://opg.optica.org/abstract.cfm?URI=optica-12-7-1020},
	doi = {10.1364/optica.561954},
	language = {en},
	number = {7},
	urldate = {2025-07-09},
	journal = {Optica},
	author = {Moille, Grégory and Shandilya, Pradyoth and Stone, Jordan and Menyuk, Curtis and Srinivasan, Kartik},
	month = jul,
	year = {2025},
	note = {Publisher: Optica Publishing Group},
	pages = {1020},
}

@article{wu_vernier_2023,
	title = {Vernier microcombs for high-frequency carrier envelope offset and repetition rate detection},
	volume = {10},
	issn = {2334-2536},
	url = {https://opg.optica.org/abstract.cfm?URI=optica-10-5-626},
	doi = {10.1364/OPTICA.486755},
	language = {en},
	number = {5},
	urldate = {2025-11-22},
	journal = {Optica},
	author = {Wu, Kaiyi and O’Malley, Nathan P. and Fatema, Saleha and Wang, Cong and Girardi, Marcello and Alshaykh, Mohammed S. and Ye, Zhichao and Leaird, Daniel E. and Qi, Minghao and Torres-Company, Victor and Weiner, Andrew M.},
	month = may,
	year = {2023},
	pages = {626},
}

@article{pfeiffer_octave-spanning_2017,
	title = {Octave-spanning dissipative {Kerr} soliton frequency combs in {Si}\_3N\_4 microresonators},
	volume = {4},
	issn = {2334-2536},
	url = {https://opg.optica.org/abstract.cfm?URI=optica-4-7-684},
	doi = {10.1364/OPTICA.4.000684},
	language = {en},
	number = {7},
	urldate = {2025-11-22},
	journal = {Optica},
	author = {Pfeiffer, Martin H. P. and Herkommer, Clemens and Liu, Junqiu and Guo, Hairun and Karpov, Maxim and Lucas, Erwan and Zervas, Michael and Kippenberg, Tobias J.},
	month = jul,
	year = {2017},
	pages = {684},
}

@article{yu_tuning_2019,
	title = {Tuning {Kerr}-{Soliton} {Frequency} {Combs} to {Atomic} {Resonances}},
	volume = {11},
	issn = {2331-7019},
	url = {https://link.aps.org/doi/10.1103/PhysRevApplied.11.044017},
	doi = {10.1103/PhysRevApplied.11.044017},
	language = {en},
	number = {4},
	urldate = {2021-10-13},
	journal = {Physical Review Applied},
	author = {Yu, Su-Peng and Briles, Travis C. and Moille, Gregory T. and Lu, Xiyuan and Diddams, Scott A. and Srinivasan, Kartik and Papp, Scott B.},
	month = apr,
	year = {2019},
	pages = {044017},
}

@article{zang_foundry_2024,
	title = {Foundry manufacturing of octave-spanning microcombs},
	volume = {49},
	issn = {0146-9592, 1539-4794},
	url = {https://opg.optica.org/abstract.cfm?URI=ol-49-18-5143},
	doi = {10.1364/OL.527540},
	language = {en},
	number = {18},
	urldate = {2025-11-22},
	journal = {Optics Letters},
	author = {Zang, Jizhao and Liu, Haixin and Briles, Travis C. and Papp, Scott B.},
	month = sep,
	year = {2024},
	pages = {5143},
}

@article{weng_dual-mode_2022,
	title = {Dual-mode microresonators as straightforward access to octave-spanning dissipative {Kerr} solitons},
	volume = {7},
	issn = {2378-0967},
	url = {https://pubs.aip.org/app/article/7/6/066103/2835160/Dual-mode-microresonators-as-straightforward},
	doi = {10.1063/5.0089036},
	language = {en},
	number = {6},
	urldate = {2025-11-22},
	journal = {APL Photonics},
	author = {Weng, Haizhong and Afridi, Adnan Ali and Li, Jing and McDermott, Michael and Tu, Huilan and Barry, Liam P. and Lu, Qiaoyin and Guo, Weihua and Donegan, John F.},
	month = jun,
	year = {2022},
	pages = {066103},
}

@article{okawachi_octave-spanning_2011,
	title = {Octave-spanning frequency comb generation in a silicon nitride chip},
	volume = {36},
	copyright = {https://doi.org/10.1364/OA\_License\_v1\#VOR},
	issn = {0146-9592, 1539-4794},
	url = {https://opg.optica.org/abstract.cfm?URI=ol-36-17-3398},
	doi = {10.1364/OL.36.003398},
	language = {en},
	number = {17},
	urldate = {2025-11-22},
	journal = {Optics Letters},
	author = {Okawachi, Yoshitomo and Saha, Kasturi and Levy, Jacob S. and Wen, Y. Henry and Lipson, Michal and Gaeta, Alexander L.},
	month = sep,
	year = {2011},
	pages = {3398},
}

@article{brasch_photonic_2016,
	title = {Photonic chip–based optical frequency comb using soliton {Cherenkov} radiation},
	volume = {351},
	issn = {0036-8075, 1095-9203},
	url = {https://www.science.org/doi/10.1126/science.aad4811},
	doi = {10.1126/science.aad4811},
	language = {en},
	number = {6271},
	urldate = {2025-11-22},
	journal = {Science},
	author = {Brasch, V. and Geiselmann, M. and Herr, T. and Lihachev, G. and Pfeiffer, M. H. P. and Gorodetsky, M. L. and Kippenberg, T. J.},
	month = jan,
	year = {2016},
	pages = {357--360},
}

@article{moille_tailoring_2021,
	title = {Tailoring broadband {Kerr} soliton microcombs via post-fabrication tuning of the geometric dispersion},
	volume = {119},
	issn = {0003-6951, 1077-3118},
	url = {https://aip.scitation.org/doi/10.1063/5.0061238},
	doi = {10.1063/5.0061238},
	language = {en},
	number = {12},
	urldate = {2021-11-10},
	journal = {Applied Physics Letters},
	author = {Moille, Gregory and Westly, Daron and Orji, Ndubuisi George and Srinivasan, Kartik},
	month = sep,
	year = {2021},
	pages = {121103},
}

@article{Luke:15,
author = {Kevin Luke and Yoshitomo Okawachi and Michael R. E. Lamont and Alexander L. Gaeta and Michal Lipson},
journal = {Opt. Lett.},
number = {21},
pages = {4823--4826},
publisher = {Optica Publishing Group},
title = {Broadband mid-infrared frequency comb generation in a Si3N4 microresonator},
volume = {40},
month = {Nov},
year = {2015},
url = {https://opg.optica.org/ol/abstract.cfm?URI=ol-40-21-4823},
doi = {10.1364/OL.40.004823},
}

\end{document}